\documentclass[apl, twocolumn,tightenlines,superscriptaddress]{revtex4-2}

\usepackage{amsthm}
\usepackage{amsmath}
\usepackage{bbm, dsfont}
\usepackage{graphicx}
\usepackage[usenames,dvipsnames]{color}
\usepackage[colorlinks=true,citecolor=magenta,urlcolor=blue]{hyperref}
\usepackage[table]{xcolor}
\usepackage{colortbl}
\usepackage{color}
\usepackage{hhline}
\usepackage{tabu}
\usepackage{epstopdf}
\usepackage{booktabs}
\usepackage[normalem]{ulem}
\usepackage{siunitx}
\usepackage{enumitem}
\usepackage{braket}
\usepackage{siunitx}
\usepackage{textcomp}
\usepackage{booktabs}
\usepackage[textwidth=3cm, textsize=footnotesize]{todonotes}
\usepackage[font=small,skip=3pt]{caption}
\usepackage{bbold}
\usepackage[percent]{overpic}
\usepackage{makecell}
\usepackage{multirow}
\usepackage{float}
\usepackage{textcomp}
\usepackage{blindtext}
\usepackage{booktabs}

\newcommand{\affiliationGap}{Group of Applied Physics, Rue de l'Ecole-de-Médecine 20, CH-1211 Genève 4, Switzerland}
\newcommand{\affiliationIDQ}{ID Quantique SA, Rue Eugène-Marziano 25, CH-1227 Acacias - Genève Switzerland}
\newcommand{\affiliationHSLU}{Lucerne School of Computer Science and Information Technology, Suurstoffi 1, CH-6343 Rotkreuz, Switzerland}

\graphicspath{{./Figures/}}
\newcolumntype{C}[1]{ >{ \centering\arraybackslash}p{#1}}

\DeclareSIUnit\permille{\text{\textperthousand}}
\DeclareSIUnit{\belmilliwatt}{Bm}
\DeclareSIUnit{\dBm}{\deci\belmilliwatt}
\DeclareSIUnit{\bps}{bps}
\DeclareSIUnit{\bit}{b}
\DeclareSIUnit{\byte}{B}
\DeclareSIUnit{\cps}{cps}

\newcommand{\pza}{p_\text{Z,A}}
\newcommand{\pzb}{p_\text{Z,B}}

\newcommand{\qubitRepRate}{\SI{2.5}{\giga\hertz}}
\newcommand{\laserFWHM}{\SI{45}{\pico\second}}

\newcommand{\fiberDistanceShort}{10.0}
\newcommand{\fiberDistanceLong}{102.4}

\newcommand{\fiberAttenuationShort}{1.58}
\newcommand{\fiberAttenuationLong}{16.34}

\newcommand{\muZeroShort}{0.49}
\newcommand{\muZeroLong}{0.46}

\newcommand{\muOneShort}{0.22}
\newcommand{\muOneLong}{0.20}

\newcommand{\pmuZeroShort}{0.74}
\newcommand{\pmuZeroLong}{0.79}

\newcommand{\pzaShort}{0.65}
\newcommand{\pzaLong}{0.66}

\newcommand{\pzbShort}{0.99}
\newcommand{\pzbLong}{0.99}

\newcommand{\siftedKRShort}{159.4}
\newcommand{\siftedKRLong}{7.8}

\newcommand{\phizShort}{0.8}
\newcommand{\phizLong}{1.0}

\newcommand{\qberzShort}{0.4}
\newcommand{\qberzLong}{0.3}

\newcommand{\SKRShort}{64}
\newcommand{\SKRLong}{3.0}

\newcommand{\efficiencyX}{0.85} 

\newcommand{\jitterX}{55} 

\begin{document}

\title{Fast Single Photon Detectors and real-time Key Distillation: Enabling High Secret Key Rate QKD Systems}

\author{Fadri Gr\"unenfelder}\email{fadri.gruenenfelder@unige.ch}
\affiliation{\affiliationGap{}}
\author{Alberto Boaron}
\affiliation{\affiliationGap{}}
\author{Matthieu Perrenoud}
\affiliation{\affiliationGap{}}
\author{Giovanni V. Resta}
\affiliation{\affiliationGap{}}
\author{Davide Rusca}
\affiliation{\affiliationGap{}}
\author{Claudio Barreiro}
\affiliation{\affiliationGap{}}
\author{Rapha\"el Houlmann}
\affiliation{\affiliationGap{}}
\author{Rebecka Sax}
\affiliation{\affiliationGap{}}
\author{Lorenzo Stasi}
\affiliation{\affiliationGap{}}
\affiliation{\affiliationIDQ{}}
\author{Sylvain El-Khoury}
\affiliation{\affiliationIDQ{}}
\author{Esther Hänggi}
\affiliation{\affiliationHSLU{}}
\author{Nico Bosshard}
\affiliation{\affiliationHSLU{}}
\author{Félix Bussières}
\affiliation{\affiliationIDQ{}}
\author{Hugo Zbinden}
\affiliation{\affiliationGap{}}

\begin{abstract}
Quantum Key Distribution has made continuous progress over the last 20 years and is now commercially available. However, the secret key rates (SKR) are still limited to a few \si{\mega \bps}. Here, we present a custom multipixel superconducting nanowire single-photon detectors and fast acquisition and real-time key distillation electronics, removing two roadblocks and allowing an increase of the SKR of more than an order of magnitude. In combination with a simple 2.5 GHz clocked time-bin quantum key distribution system, we can generate secret keys at a rate of \SI{\SKRShort}{\mega\bps} over a distance of \SI{\fiberDistanceShort}{\kilo\meter} and at a rate of \SI{\SKRLong}{\mega\bps} over a distance of \SI{\fiberDistanceLong}{\kilo\meter} with real-time key distillation. 

\end{abstract}

\maketitle


Quantum key distribution (QKD) allows the exchange of cryptographic keys at a distance without assumptions on the technological limits of a possible eavesdropper, in particular their computational power \cite{Bennett1984,Ekert1991}. In contrast, currently used public key systems rely on computationally demanding tasks \cite{Rivest1978,Koblitz1987}. While nowadays an eavesdropper is bound to use classical computers, this could change in the near future with the advent of large scale quantum computers. This renders an eavesdropper able to use powerful attacks which today's public key systems cannot withstand \cite{Shor1995}. The security of QKD, however, is solely based on the laws of quantum mechanics. Therefore, together with the One-Time-Pad \cite{Shannon1949}, private communication can be ensured even in a future where quantum computers are widely available.

Since the advent of the QKD era with the BB84 protocol \cite{Bennett1984}, a variety of other protocols have been developed \cite{Ekert1991,Lo2012,Lucamarini2018,Ralph1999}. Although the complexity and level of device-independence differ between protocols, the main goals remain the same, namely to increase the distance over which a secret key can be generated or, conversely, to maximize the secret key rate (SKR) over a certain distance. To give some context, we consider the use-case of encrypted video conferencing. The United States Federal Communications Commission recommends a download rate of \SI{6}{\mega\bps} for this application, so with One-Time-Pad encryption one needs a SKR equal to this rate per user. For more demanding applications, such as data centers, much higher SKRs are required. Recently, it was demonstrated that a single QKD link can achieve a SKR up to \SI{13.72}{\mega\bps} over a channel equivalent to \SI{10}{\kilo\meter} of single-mode fiber \cite{Yuan2018}. A proof-of-principle experiment using space division multiplexing showed that it would be possible to achieve a SKR of \SI{105.7}{\mega\bps} over a distance of \SI{7.9}{\km} by using 37 QKD transmitters and receivers with a multi-core fiber as quantum channel \cite{Bacco2019}.

In order to go to even higher secret  key rates without multiplexing, a QKD system needs to fulfill a couple of requirements:
Obviously, the transmitter must emit qubits at a high repetition rate. However, a high repetition is only useful, in particular at shorter distances, if i) the single photon detectors are able to count at high rates (with high efficiency and low timing jitter), ii) the readout and sifting electronics are able to process these rates and iii) the post-processing unit is capable of correcting the key (with low leakage) and performing privacy amplification in \textbf{real-time}. 

In this paper, we report on our efforts of improving points i) to iii). In particular, we present a custom Superconducting Nanowire Single Photon Detector (SNSPD) featuring high count rates and high efficiency. We discuss how to optimize the parameters of a QKD system for high secret key rates and demonstrate an implementation generating a SKR of more than \SI{60}{\mega \bps} over \SI{10}{\km} and \SI{2}{\mega\bps} over 100 km. We use a simplified BB84 with time-bin encoding and one decoy state clocked at 2.5 GHz (time bins of 100 ps seperated by 100 ps) \cite{Rusca2018, Rusca2018b, Boaron2018b}, but the presented principles are valid also for polarisation based schemes (see \cite{Li2022}). 


\begin{figure}
	\includegraphics[width = .9 \columnwidth]{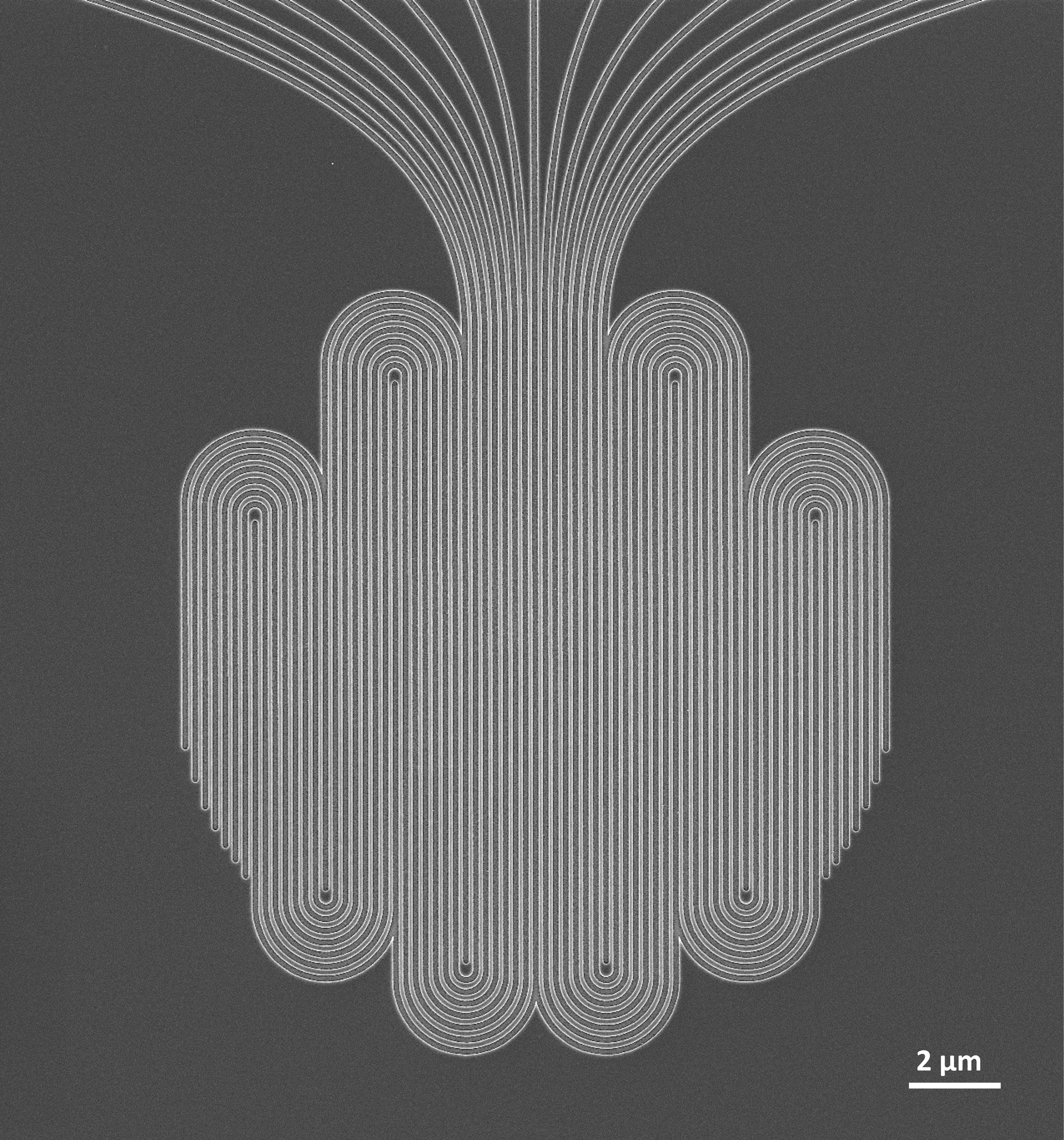}
	\caption{\label{fig:SNSPD} Image of the 14 interleaved pixels SNSPD taken with a scanning electron microscope.}
\end{figure}

Let's start with a description and characterization of the SNSPD developed in-house. We design the detector such that high efficiency, low jitter and a high maximum count rate can be achieved \textbf{simultaneously}. This is not an easy task, as the jitter of an SNSPDs tends to increase at high count rates.

\begin{figure}
	\includegraphics[width = 1.\columnwidth]{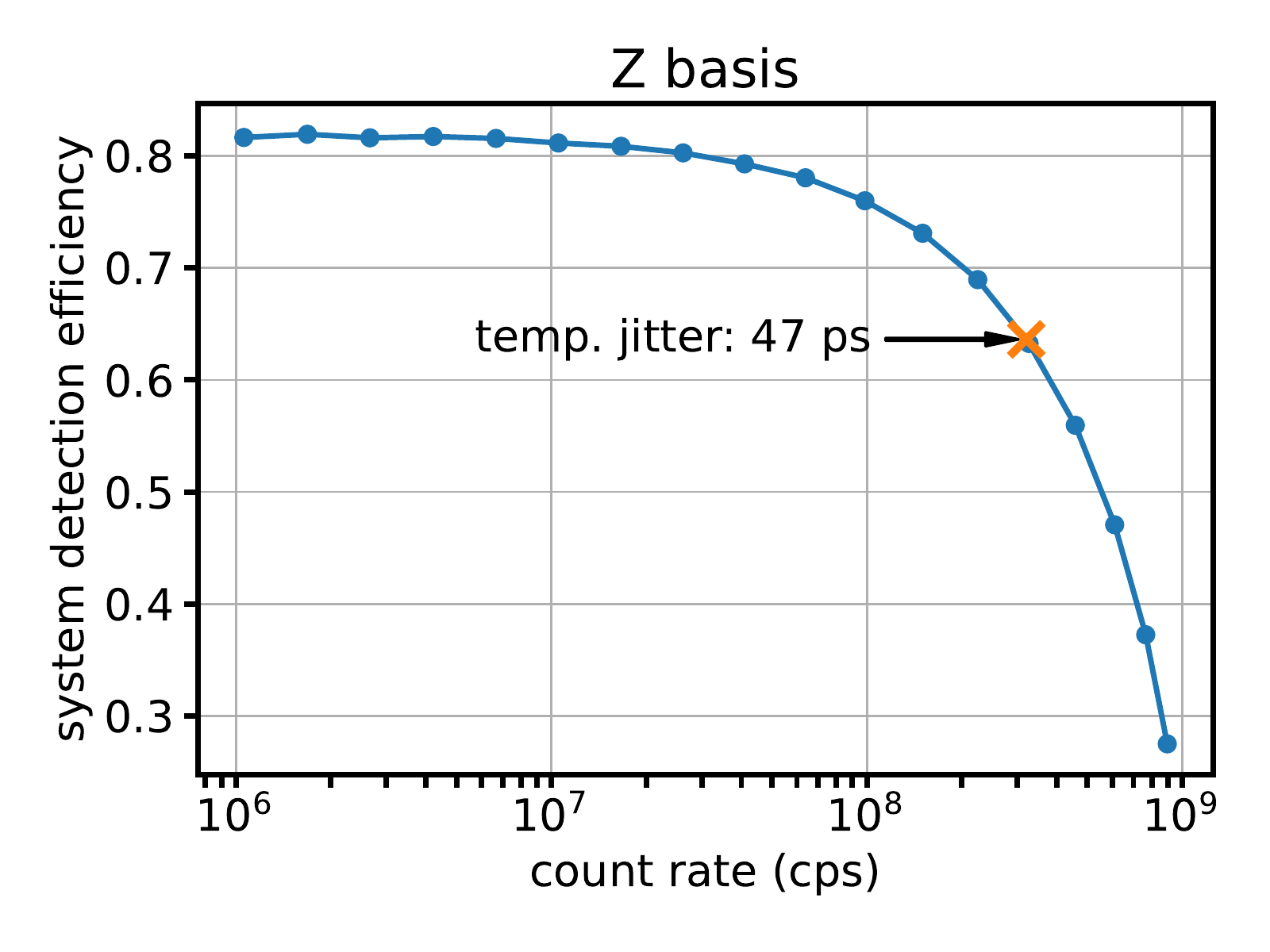}
	\caption{\label{fig:eff_vs_counts_z} Detection efficiency of the multipixel detector used in the Z basis. The insert shows the jitter measured at the count rate marked by the orange cross.}
\end{figure}

As the superconducting material for our nanowires, we use niobium-titanium nitride (NbTiN) that has been sputtered from a NbTi target in a nitrogen-rich atmosphere. The superconducting film has a thickness of around \SI{9}{\nm} and exhibits a critical temperature ($T_\text{c}$) of \SI{8.5}{\kelvin}. The detector is composed of 14 independent pixels arranged in an interleaved geometry (see \autoref{fig:SNSPD}). The number of pixels is chosen to comply with the requirements of Bob, and the generated signal is amplified at \SI{40}{\K} with a custom-made amplifier board. Thanks to the large number of pixels and the interleaved design, which guarantees an uniform illumination of the pixels, the probability that two detections occur during the recovery time on the same pixel is minimized. The detector is integrated in an optical cavity, designed to maximize photon absorption at \SI{1550}{\nm}, and exhibits a maximum system detection efficiency of 82\%, see \autoref{fig:eff_vs_counts_z}. The detector covers the same area as a conventional single-pixel SNSPD (around \SI{200}{\um^2}), thus the length of each nanowire is greatly reduced, allowing for much faster recovery time (on average, $< \SI{10}{\ns}$ to be back at full efficiency) compared to a single-meander SNSPD. The fast recovery time of each pixel directly translates into the capability to reach \si{\GHz} detection rates when reading all the 14 pixels simultaneously. However, Bob must measure the arrival time of the pulses and be sure to assign them to the correct time bin, that has a \SI{100}{\ps} duration, thus posing a stringent requirement on the jitter of the detector and of the whole read-out system.

We characterize the jitter of the 14-pixel detector at a count rate of \SI{1}{\mega\cps}. There, the average jitter for the 14 pixels is \SI{22}{\ps}, which represents a good starting point for when the count rate will be pushed to the limit. In fact, at the high rate of the QKD system, more and more photon detections occur when the bias current in the SNSPD has not yet reached its maximum value, i.e. before the current and efficiency have fully recovered, thus causing an increase of the jitter. One contribution to the jitter at high detection rates is the variation in the amplitude of the detection signal. This contribution can be minimized by using constant fraction discriminators (CFDs) instead of threshold discriminators. We designed and built CFDs which are optimized for the use with our multipixel detector. With this readout electronics we obtained a jitter below \SI{47}{\ps} and an efficiency of 64\% at a count rate of 320 Mcps.

\begin{figure*}
\includegraphics[width = 1.95\columnwidth]{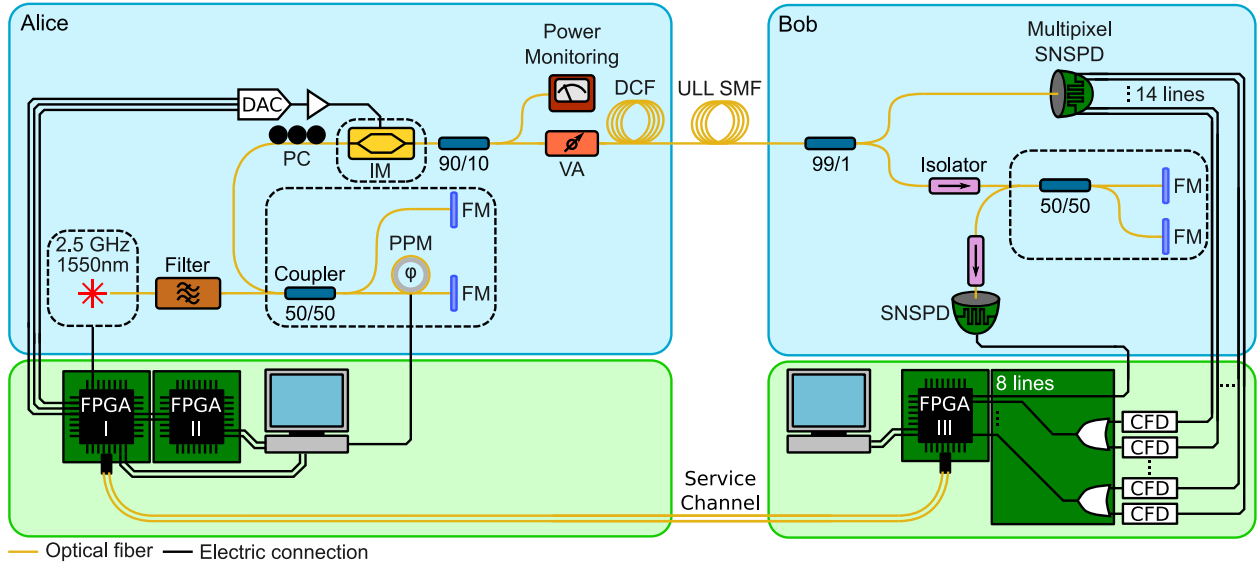}
\caption{\label{fig:setup} Schematic of the setup.  
CFD: constant fraction discriminator;
DAC: digital-to-analog converter;
DCF: dispersion-compensating fiber;
FM: Faraday mirror;
FPGA: field-programmable gate array;
IM: intensity modulator;
PC: polarization controller;
PPM: piezo-electric phase modulator; 
SNSPD: superconducting nanowire single-photon detector; 
ULL SMF: ultra-low-loss single-mode fiber;
VA: variable attenuator; The dashed boxes are temperature stabilized. FPGA I controls the state preparation, FPGA II is used for error correction and FGPA III acquires the detection events. The sifting is done between FPGAs I and III. Both Michelson interferometers exhibit an imbalance of \SI{200}{\pico\second}.
}
\end{figure*}

For error correction, we use a quasi-cyclic low-density parity check code (LDPC) with a syndrome size of $1/6$ which is implemented in the FPGAs. Bob calculates the syndrome and sends it to Alice. The resource-intensive error correction core is running on the FPGA II (Xilinx Virtex-6 LXT, see \autoref{fig:setup}) on Alice's side. One core can correct up to \SI{110}{\mega\cps}. By simply running two cores in parallel on the same FPGA, we achieve a throughput of \SI{220}{\mega\cps}, which is high enough for our experiment. The privacy amplification is implemented on a consumer-type computer. It receives the sifted key via a Generation 2 PCIe x4 connector (maximum throughput of \SI{4}{\giga\byte}) from the FPGA. It runs on a consumer-type graphics processing unit (RTX 2070 Super Ventus OC) and has a maximum throughput of \SI{3.4}{\giga\bps}. The block size of the algorithm is $2^{27}~\text{bit} \approx 134~\text{Mbit}$ and the secrecy parameter we used is $10^{-15}$ (for more details on the extraction see \cite{Bosshard2021}).



The ultra-fast SNSPD and the post-processing error correction are implemented in our QKD system, as shown in  \autoref{fig:setup}.

\begin{figure}
	\includegraphics[width = 1.\columnwidth]{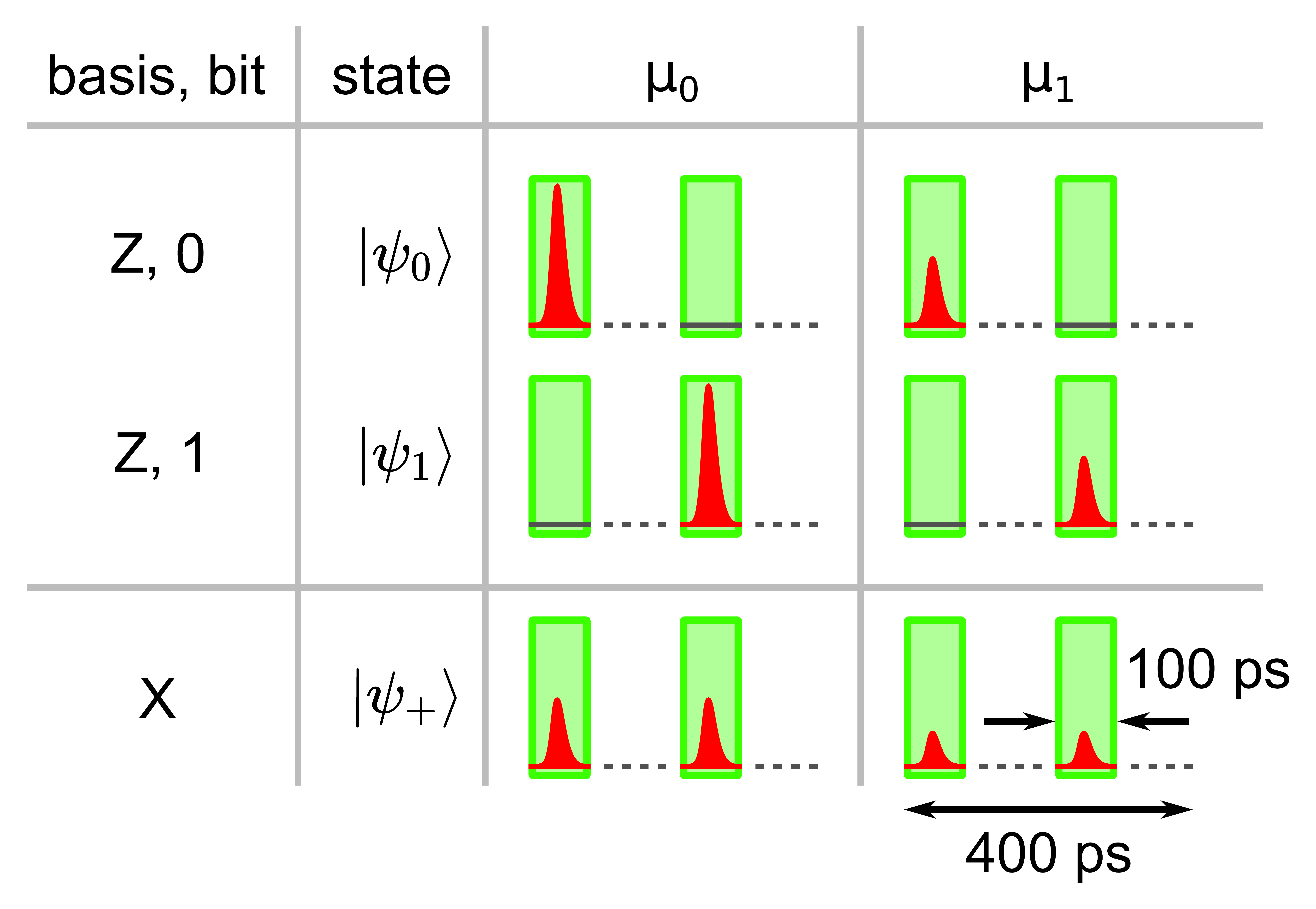}
	\caption{\label{fig:states} States prepared by Alice. Each state consists of two time bins. The two pulses of one state have a fixed phase relation, but pulses of different states have a random phase relation. Alice chooses the mean photon number $\mu_0$ or $\mu_1$ for each state at random. The green boxes are the detection time windows of Bob, each with a duration of \SI{100}{\pico\second}.
	}
\end{figure}

We use the simplified BB84 with time-bin encoding and one decoy state \cite{Rusca2018, Rusca2018b, Boaron2018b}. Alice prepares the states shown in \autoref{fig:states}. The two states $\ket{0}$ and $\ket{1}$ form the Z basis and $\ket{+}$ the X basis. She chooses the basis at random with probabilities $\pza$ for the Z basis and $1 - \pza$ for the X basis. In case she chooses the Z basis, she picks either  $\ket{0}$ or $\ket{1}$ with equal probability. Additionally, she chooses at random between two mean photon numbers $\mu_0$ and $\mu_1$ with probabilities $p_{\mu_0}$ and $p_{\mu_1}$. Bob picks a measurement basis at random with probabilities $\pzb$ for the Z basis and $1 - \pzb$ for the X basis. The secret key rate is generated from the correlations in the Z basis and they use a privacy amplification block size of $2^{27}~\text{bit} \approx 134~\text{Mbit}$. The X basis is used to find an upper bound on the phase error rate via the decoy method \cite{Rusca2018b}.

 A distributed feedback InGaAsP/InP multi-quantum well laser diode is used to create a train of phase-randomized pulses with a full width at half maximum (FWHM) of \laserFWHM{} and at a rate of \qubitRepRate{}. Then the pulses pass an imbalanced Michelson interferometer with a time difference of \SI{200}{\pico\second} between the two arms. The states are encoded using an intensity modulator (IM). The three states with the two mean photon levels required for this protocol are shown in \autoref{fig:states}. In the Z basis, we have a pulse either in the early or in the late time bin. The state in the X basis carries pulses in both time bins, but with half the intensity than the pulses in the Z basis.  Alice chooses the Z basis with a probability $\pza$ depending on the distance. In any case, the value of $\pza$ is well above 0.5. These two pulses have a fixed phase relation, while in between the states, the phase is randomized due to the gain-switching of the laser. As a quantum channel serves a ULL single mode fiber. Its dispersion is pre-compensated by dispersion-compensating fiber. 

At the other end of the channel, at Bob's, the basis is selected passively with the help of a fibre coupler. The optimal probability $\pzb$ is close to unity. This means that the sifting efficiency is significantly higher than in the standard BB84. In the Z basis, Bob measures the time of arrival of the signal with the multipixel detector described above. In the X basis, the pulses pass through a Michelson interferometer with the same delay as the one of Alice. Here, the requirement on the detector is less stringent due to the high bias of the basis choice towards the Z basis. We choose a MoSi SNSPD with a parallel design which exhibits a timing jitter below \SI{\jitterX}{\pico\second} and an efficiency of \efficiencyX{} at a count rate of \SI{2}{\mega\cps}. 

The states prepared by Alice and the events measured by Bob are registered by FPGA I and III (Xilinx Kintex-7 FPGA KC705 Evaluation Kit). The outputs of the detectors are interfaced to the FPGA with an in-house made card. This card can delay the 14 channels of the multipixel detector and the channel of the X basis detector individually, allowing us to synchronize them. Further, the card combines the 14 channels of the multipixel detectors into 7 channels with OR-gates. At very high count rates, the combining of channels will mask some detections. By comparing the count rate of the QKD system with the count rate measured with time-to-digital converters, we found that due to the OR-gate masking, we loose 2.8\% of the counts at \SI{320}{\mega\cps}.

The two FPGAs communicate directly via a service channel to perform sifting and error correction in real-time. 

\begin{figure}
	\includegraphics[width = 1.\columnwidth]{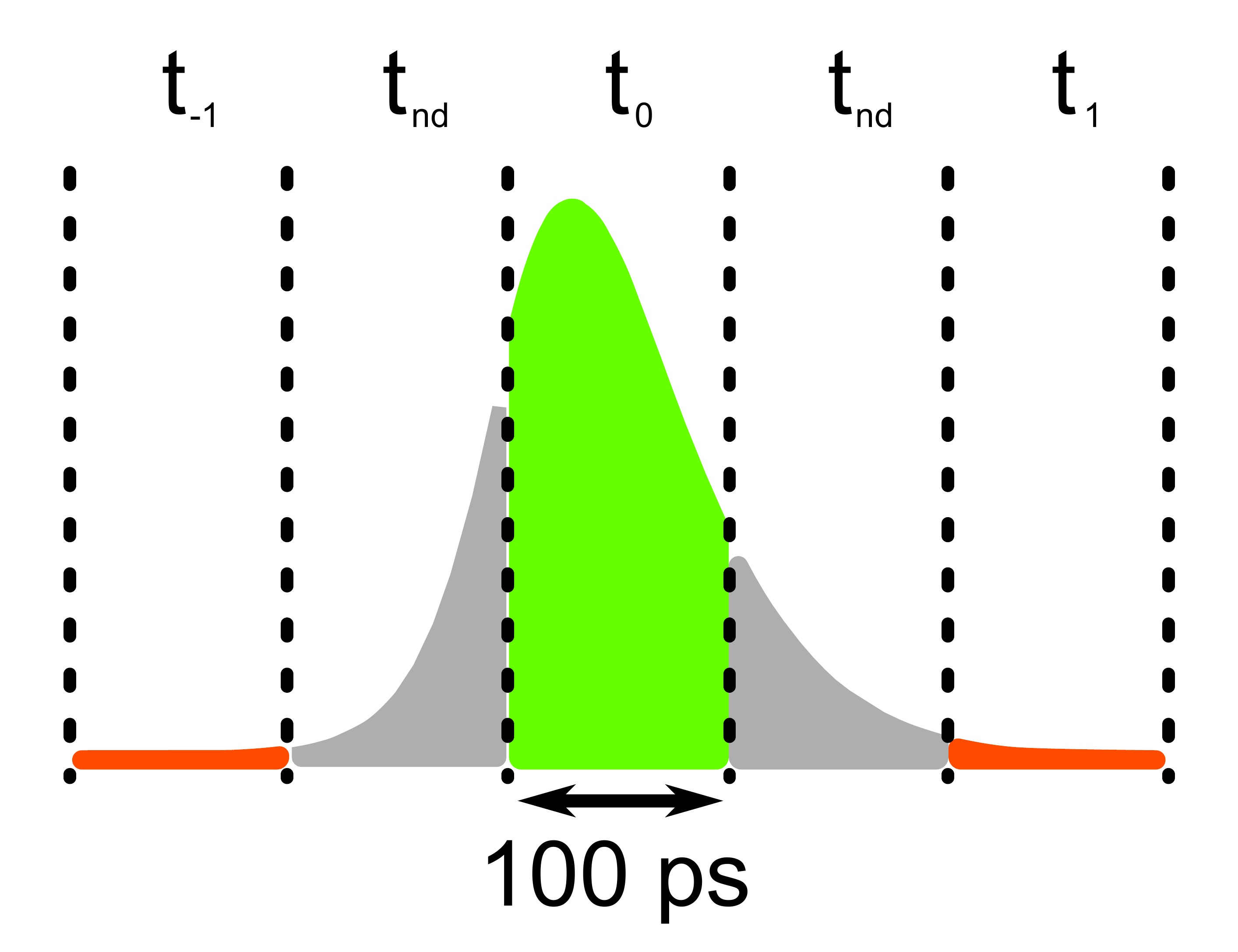}
	\caption{\label{fig:jitter} THe time bin division is represented as $t_0$ correct time-bin of arrival, $t_{-1}$ and $t_1$ are respectively the previous time-bin and the following one and $t_{nd}$ represents the time-bins discarded to lower the QBER. In this scenario the effect of the jitter is translated both in possible loss (pulse component in the $t_{nd}$ bins) or possible errors (pulse component in the $t_{-1}$ or $t_1$ bins).
	}
\end{figure}

We performed secret key exchanges through optical fibers with lengths of \SI{\fiberDistanceShort}{\kilo\meter} and \SI{\fiberDistanceLong}{\kilo\meter} for typically half an hour. The results over a privacy amplification block (for the two distances) are shown in \autoref{tab:results_skr}, together with the relevant parameters. The mean photon number of the signal and decoy states and the probabilities to choose the Z basis at Alice and Bob were obtained by numerical optimization for each distance. We manage to exchange secret keys at a rate of \SI{\SKRShort}{\mega\bps} over a distance of \SI{\fiberDistanceShort}{\kilo\meter} and at a rate of \SI{\SKRLong}{\mega\bps} over a distance of \SI{\fiberDistanceLong}{\kilo\meter}.

\begin{table*}[ht]
	\centering
	\begin{tabular}{ c c c c c c c c  c c c}\toprule
		fiber length  & att.  & $\mu_0$ & $\mu_1$ & $p_{\mu_0}$ & $\pza$ & $\pzb$ & $R_\text{sift}$ & $\phi_\text{Z}$ & $Q_\text{Z}$ & SKR  \\
		(km) & (dB) &  &  & & &  &  (Mbps) & (\%) & (\%) & (Mbps)  \\
	\midrule
		\fiberDistanceShort{} & \fiberAttenuationShort{} & \muZeroShort{} & \muOneShort{} & \pmuZeroShort{} & \pzaShort{} & \pzbShort{} & \siftedKRShort{} & \phizShort{} & \qberzShort{} & \SKRShort{} 
		\\
		\fiberDistanceLong{} & \fiberAttenuationLong{} & \muZeroLong{} & \muOneLong{} & \pmuZeroLong{} & \pzaLong{} & \pzbLong{} & \siftedKRLong{} & \phizLong{} & \qberzLong{} & \SKRLong{} 
		\\
		\bottomrule
	\end{tabular}
	\caption{\label{tab:results_skr} Measured secret key rate (SKR) and corresponding experimental parameters. The variables $\mu_0$ and $\mu_1$ stand for the mean photon number of the signal and decoy states, $p_{\mu_0}$ and $p_{\mu_1}$ are the corresponding probabilities to choose these values, $\pza$ and $\pzb$ are the probabilities of Alice and Bob to choose the Z basis,  $R_\text{sift}$ is the sifted key rate, $\phi_\text{Z}$ is the phase error rate and $Q_\text{Z}$ is the QBER Z.}
\end{table*} 


While these are best of class results, there is still some room for improvements. Indeed, our QKD scheme was designed to be simple and suitable for commercial device. To push the system to its limits in terms of maximum secret key some adaptations could be made.

Due to our high repetition rate and consequently small time bins we loose the detections that fall outside the time bins (see \autoref{fig:jitter}). Whereas further reducing the timing jitter is not simple, we could just double the multi-pixel detectors at Bob. The advantage would be twofold; the detection rate will be halved on each detector, leading to an almost 10\% increase in detection efficiency, see \autoref{fig:eff_vs_counts_z}, and a decrease in jitter. 

Our protocols allows for only one detector in the monitoring basis (projection in only one eigenstate of the X basis). In order to guarantee the security in this configuration we also need to monitor events where Alice used the Z-basis and Bob measures in the X-basis and vice-versa. Moreover, we record also events depending on which state was sent previously (some detection depend on two subsequent pulses, see \cite{Rusca2018b} for details). This forces us to chose, in the finite key scenario, a lower $\pza$ which lowers the possible achievable sifted key rate. 

Finally, the error correction is still not optimal. The used LDPC implementation has a leakage of 17\% of the sifted key rate at a QBER of 0.5\%. This is much higher than the Shannon limit of 5\% of leakage. Rate-adaptive LDPC codes could help minimizing the leakage\cite{Elkouss2010,Kiktenko2017}, but the corresponding studies do not give information about the leakage at very low QBER. Another solution would be to implement cascade (see \cite{Mao2022}) which would allow to approach the Shannon limit, in fact the state of the art allows for an efficiency of 1.038 and more than 500 MHz of throughput.

Implementing these improvements would allow us to achieve about \SI{140}{\mega\bps} at 10 km (under the condition that the other parameters stay the same).

In conclusion, we demonstrated secret key rates up to \SI{\SKRShort}{\mega\bps} over a distance of \SI{\fiberDistanceShort}{\kilo\meter}. This achievement was possible thanks to a QKD system working at a high repetition rate of \qubitRepRate, coupled with our custom SNSPDs and readout electronics which allow us to detect with low jitter and high efficiency at a high count rate. This result paves the way for secret key-demanding applications like real-time one-time-pad secured video encryption in a metropolitan area.

\section*{Acknowledgements}

We acknowledge financial support from the European Quantum Flagship project openQKD and the Swiss NCCR QSIT and from the SNSF Practice-to-Science Grant No 199084.

\section*{Data Availability}

The data that support the findings of this study are available from the corresponding author upon reasonable request.

\section*{Code Availability}

The computer code that support the findings of this study are available from the corresponding author upon reasonable request.

\bibliography{main}

\begin{thebibliography}{19}%
\makeatletter
\providecommand \@ifxundefined [1]{%
 \@ifx{#1\undefined}
}%
\providecommand \@ifnum [1]{%
 \ifnum #1\expandafter \@firstoftwo
 \else \expandafter \@secondoftwo
 \fi
}%
\providecommand \@ifx [1]{%
 \ifx #1\expandafter \@firstoftwo
 \else \expandafter \@secondoftwo
 \fi
}%
\providecommand \natexlab [1]{#1}%
\providecommand \enquote  [1]{``#1''}%
\providecommand \bibnamefont  [1]{#1}%
\providecommand \bibfnamefont [1]{#1}%
\providecommand \citenamefont [1]{#1}%
\providecommand \href@noop [0]{\@secondoftwo}%
\providecommand \href [0]{\begingroup \@sanitize@url \@href}%
\providecommand \@href[1]{\@@startlink{#1}\@@href}%
\providecommand \@@href[1]{\endgroup#1\@@endlink}%
\providecommand \@sanitize@url [0]{\catcode `\\12\catcode `\$12\catcode
  `\&12\catcode `\#12\catcode `\^12\catcode `\_12\catcode `\%12\relax}%
\providecommand \@@startlink[1]{}%
\providecommand \@@endlink[0]{}%
\providecommand \url  [0]{\begingroup\@sanitize@url \@url }%
\providecommand \@url [1]{\endgroup\@href {#1}{\urlprefix }}%
\providecommand \urlprefix  [0]{URL }%
\providecommand \Eprint [0]{\href }%
\providecommand \doibase [0]{https://doi.org/}%
\providecommand \selectlanguage [0]{\@gobble}%
\providecommand \bibinfo  [0]{\@secondoftwo}%
\providecommand \bibfield  [0]{\@secondoftwo}%
\providecommand \translation [1]{[#1]}%
\providecommand \BibitemOpen [0]{}%
\providecommand \bibitemStop [0]{}%
\providecommand \bibitemNoStop [0]{.\EOS\space}%
\providecommand \EOS [0]{\spacefactor3000\relax}%
\providecommand \BibitemShut  [1]{\csname bibitem#1\endcsname}%
\let\auto@bib@innerbib\@empty
\bibitem [{\citenamefont {Bennett}\ and\ \citenamefont
  {Brassard}(1984)}]{Bennett1984}%
  \BibitemOpen
  \bibfield  {author} {\bibinfo {author} {\bibfnamefont {C.~H.}\ \bibnamefont
  {Bennett}}\ and\ \bibinfo {author} {\bibfnamefont {G.}~\bibnamefont
  {Brassard}},\ }\bibfield  {title} {\bibinfo {title} {Quantum cryptography:
  Public key distribution and coin tossing},\ }in\ \href@noop {} {\emph
  {\bibinfo {booktitle} {International Conference on Computers, Systems \&
  Signal Processing, Bangalore, India, Dec 9-12, 1984}}}\ (\bibinfo {year}
  {1984})\ pp.\ \bibinfo {pages} {175--179}\BibitemShut {NoStop}%
\bibitem [{\citenamefont {Ekert}(1991)}]{Ekert1991}%
  \BibitemOpen
  \bibfield  {author} {\bibinfo {author} {\bibfnamefont {A.~K.}\ \bibnamefont
  {Ekert}},\ }\bibfield  {title} {\bibinfo {title} {Quantum cryptography based
  on bell's theorem},\ }\href {https://doi.org/10.1103/PhysRevLett.67.661}
  {\bibfield  {journal} {\bibinfo  {journal} {Phys. Rev. Lett.}\ }\textbf
  {\bibinfo {volume} {67}},\ \bibinfo {pages} {661} (\bibinfo {year}
  {1991})}\BibitemShut {NoStop}%
\bibitem [{\citenamefont {Rivest}\ \emph {et~al.}(1978)\citenamefont {Rivest},
  \citenamefont {Shamir},\ and\ \citenamefont {Adleman}}]{Rivest1978}%
  \BibitemOpen
  \bibfield  {author} {\bibinfo {author} {\bibfnamefont {R.~L.}\ \bibnamefont
  {Rivest}}, \bibinfo {author} {\bibfnamefont {A.}~\bibnamefont {Shamir}},\
  and\ \bibinfo {author} {\bibfnamefont {L.}~\bibnamefont {Adleman}},\
  }\bibfield  {title} {\bibinfo {title} {A method for obtaining digital
  signatures and public-key cryptosystems},\ }\href
  {https://doi.org/10.1145/359340.359342} {\bibfield  {journal} {\bibinfo
  {journal} {Communications of the {ACM}}\ }\textbf {\bibinfo {volume} {21}},\
  \bibinfo {pages} {120} (\bibinfo {year} {1978})}\BibitemShut {NoStop}%
\bibitem [{\citenamefont {Koblitz}(1987)}]{Koblitz1987}%
  \BibitemOpen
  \bibfield  {author} {\bibinfo {author} {\bibfnamefont {N.}~\bibnamefont
  {Koblitz}},\ }\bibfield  {title} {\bibinfo {title} {Elliptic curve
  cryptosystems},\ }\href {https://doi.org/10.1090/s0025-5718-1987-0866109-5}
  {\bibfield  {journal} {\bibinfo  {journal} {Mathematics of Computation}\
  }\textbf {\bibinfo {volume} {48}},\ \bibinfo {pages} {203} (\bibinfo {year}
  {1987})}\BibitemShut {NoStop}%
\bibitem [{\citenamefont {Shor}(1995)}]{Shor1995}%
  \BibitemOpen
  \bibfield  {author} {\bibinfo {author} {\bibfnamefont {P.~W.}\ \bibnamefont
  {Shor}},\ }\bibfield  {title} {\bibinfo {title} {Polynomial-time algorithms
  for prime factorization and discrete logarithms on a quantum computer}\
  }\href {https://doi.org/10.1137/S0097539795293172}
  {10.1137/S0097539795293172} (\bibinfo {year} {1995})\BibitemShut {NoStop}%
\bibitem [{\citenamefont {Shannon}(1949)}]{Shannon1949}%
  \BibitemOpen
  \bibfield  {author} {\bibinfo {author} {\bibfnamefont {C.~E.}\ \bibnamefont
  {Shannon}},\ }\bibfield  {title} {\bibinfo {title} {Communication theory of
  secrecy systems},\ }\href
  {https://doi.org/10.1002/j.1538-7305.1949.tb00928.x} {\bibfield  {journal}
  {\bibinfo  {journal} {Bell System Technical Journal}\ }\textbf {\bibinfo
  {volume} {28}},\ \bibinfo {pages} {656} (\bibinfo {year} {1949})}\BibitemShut
  {NoStop}%
\bibitem [{\citenamefont {Lo}\ \emph {et~al.}(2012)\citenamefont {Lo},
  \citenamefont {Curty},\ and\ \citenamefont {Qi}}]{Lo2012}%
  \BibitemOpen
  \bibfield  {author} {\bibinfo {author} {\bibfnamefont {H.-K.}\ \bibnamefont
  {Lo}}, \bibinfo {author} {\bibfnamefont {M.}~\bibnamefont {Curty}},\ and\
  \bibinfo {author} {\bibfnamefont {B.}~\bibnamefont {Qi}},\ }\bibfield
  {title} {\bibinfo {title} {Measurement-device-independent quantum key
  distribution},\ }\href {https://doi.org/10.1103/PhysRevLett.108.130503}
  {\bibfield  {journal} {\bibinfo  {journal} {Phys. Rev. Lett.}\ }\textbf
  {\bibinfo {volume} {108}},\ \bibinfo {pages} {130503} (\bibinfo {year}
  {2012})}\BibitemShut {NoStop}%
\bibitem [{\citenamefont {Lucamarini}\ \emph {et~al.}(2018)\citenamefont
  {Lucamarini}, \citenamefont {Yuan}, \citenamefont {Dynes},\ and\
  \citenamefont {Shields}}]{Lucamarini2018}%
  \BibitemOpen
  \bibfield  {author} {\bibinfo {author} {\bibfnamefont {M.}~\bibnamefont
  {Lucamarini}}, \bibinfo {author} {\bibfnamefont {Z.~L.}\ \bibnamefont
  {Yuan}}, \bibinfo {author} {\bibfnamefont {J.~F.}\ \bibnamefont {Dynes}},\
  and\ \bibinfo {author} {\bibfnamefont {A.~J.}\ \bibnamefont {Shields}},\
  }\bibfield  {title} {\bibinfo {title} {Overcoming the
  rate{\textendash}distance limit of quantum key distribution without quantum
  repeaters},\ }\href {https://doi.org/10.1038/s41586-018-0066-6} {\bibfield
  {journal} {\bibinfo  {journal} {Nature}\ }\textbf {\bibinfo {volume} {557}},\
  \bibinfo {pages} {400} (\bibinfo {year} {2018})}\BibitemShut {NoStop}%
\bibitem [{\citenamefont {Ralph}(1999)}]{Ralph1999}%
  \BibitemOpen
  \bibfield  {author} {\bibinfo {author} {\bibfnamefont {T.~C.}\ \bibnamefont
  {Ralph}},\ }\bibfield  {title} {\bibinfo {title} {Continuous variable quantum
  cryptography},\ }\bibfield  {journal} {\bibinfo  {journal} {Physical Review
  A}\ }\textbf {\bibinfo {volume} {61}},\ \href
  {https://doi.org/10.1103/physreva.61.010303} {10.1103/physreva.61.010303}
  (\bibinfo {year} {1999})\BibitemShut {NoStop}%
\bibitem [{\citenamefont {Yuan}\ \emph {et~al.}(2018)\citenamefont {Yuan},
  \citenamefont {Murakami}, \citenamefont {Kujiraoka}, \citenamefont
  {Lucamarini}, \citenamefont {Tanizawa}, \citenamefont {Sato}, \citenamefont
  {Shields}, \citenamefont {Plews}, \citenamefont {Takahashi}, \citenamefont
  {Doi}, \citenamefont {Tam}, \citenamefont {Sharpe}, \citenamefont {Dixon},
  \citenamefont {Lavelle},\ and\ \citenamefont {Dynes}}]{Yuan2018}%
  \BibitemOpen
  \bibfield  {author} {\bibinfo {author} {\bibfnamefont {Z.}~\bibnamefont
  {Yuan}}, \bibinfo {author} {\bibfnamefont {A.}~\bibnamefont {Murakami}},
  \bibinfo {author} {\bibfnamefont {M.}~\bibnamefont {Kujiraoka}}, \bibinfo
  {author} {\bibfnamefont {M.}~\bibnamefont {Lucamarini}}, \bibinfo {author}
  {\bibfnamefont {Y.}~\bibnamefont {Tanizawa}}, \bibinfo {author}
  {\bibfnamefont {H.}~\bibnamefont {Sato}}, \bibinfo {author} {\bibfnamefont
  {A.~J.}\ \bibnamefont {Shields}}, \bibinfo {author} {\bibfnamefont
  {A.}~\bibnamefont {Plews}}, \bibinfo {author} {\bibfnamefont
  {R.}~\bibnamefont {Takahashi}}, \bibinfo {author} {\bibfnamefont
  {K.}~\bibnamefont {Doi}}, \bibinfo {author} {\bibfnamefont {W.}~\bibnamefont
  {Tam}}, \bibinfo {author} {\bibfnamefont {A.~W.}\ \bibnamefont {Sharpe}},
  \bibinfo {author} {\bibfnamefont {A.~R.}\ \bibnamefont {Dixon}}, \bibinfo
  {author} {\bibfnamefont {E.}~\bibnamefont {Lavelle}},\ and\ \bibinfo {author}
  {\bibfnamefont {J.~F.}\ \bibnamefont {Dynes}},\ }\bibfield  {title} {\bibinfo
  {title} {10-mb/s quantum key distribution},\ }\href
  {https://doi.org/10.1109/jlt.2018.2843136} {\bibfield  {journal} {\bibinfo
  {journal} {Journal of Lightwave Technology}\ }\textbf {\bibinfo {volume}
  {36}},\ \bibinfo {pages} {3427} (\bibinfo {year} {2018})}\BibitemShut
  {NoStop}%
\bibitem [{\citenamefont {Bacco}\ \emph {et~al.}(2019)\citenamefont {Bacco},
  \citenamefont {Lio}, \citenamefont {Cozzolino}, \citenamefont {Ros},
  \citenamefont {Guo}, \citenamefont {Ding}, \citenamefont {Sasaki},
  \citenamefont {Aikawa}, \citenamefont {Miki}, \citenamefont {Terai},
  \citenamefont {Yamashita}, \citenamefont {Neergaard-Nielsen}, \citenamefont
  {Galili}, \citenamefont {Rottwitt}, \citenamefont {Andersen}, \citenamefont
  {Morioka},\ and\ \citenamefont {Oxenl{\o}we}}]{Bacco2019}%
  \BibitemOpen
  \bibfield  {author} {\bibinfo {author} {\bibfnamefont {D.}~\bibnamefont
  {Bacco}}, \bibinfo {author} {\bibfnamefont {B.~D.}\ \bibnamefont {Lio}},
  \bibinfo {author} {\bibfnamefont {D.}~\bibnamefont {Cozzolino}}, \bibinfo
  {author} {\bibfnamefont {F.~D.}\ \bibnamefont {Ros}}, \bibinfo {author}
  {\bibfnamefont {X.}~\bibnamefont {Guo}}, \bibinfo {author} {\bibfnamefont
  {Y.}~\bibnamefont {Ding}}, \bibinfo {author} {\bibfnamefont {Y.}~\bibnamefont
  {Sasaki}}, \bibinfo {author} {\bibfnamefont {K.}~\bibnamefont {Aikawa}},
  \bibinfo {author} {\bibfnamefont {S.}~\bibnamefont {Miki}}, \bibinfo {author}
  {\bibfnamefont {H.}~\bibnamefont {Terai}}, \bibinfo {author} {\bibfnamefont
  {T.}~\bibnamefont {Yamashita}}, \bibinfo {author} {\bibfnamefont {J.~S.}\
  \bibnamefont {Neergaard-Nielsen}}, \bibinfo {author} {\bibfnamefont
  {M.}~\bibnamefont {Galili}}, \bibinfo {author} {\bibfnamefont
  {K.}~\bibnamefont {Rottwitt}}, \bibinfo {author} {\bibfnamefont {U.~L.}\
  \bibnamefont {Andersen}}, \bibinfo {author} {\bibfnamefont {T.}~\bibnamefont
  {Morioka}},\ and\ \bibinfo {author} {\bibfnamefont {L.~K.}\ \bibnamefont
  {Oxenl{\o}we}},\ }\bibfield  {title} {\bibinfo {title} {Boosting the secret
  key rate in a shared quantum and classical fibre communication system},\
  }\bibfield  {journal} {\bibinfo  {journal} {Communications Physics}\ }\textbf
  {\bibinfo {volume} {2}},\ \href {https://doi.org/10.1038/s42005-019-0238-1}
  {10.1038/s42005-019-0238-1} (\bibinfo {year} {2019})\BibitemShut {NoStop}%
\bibitem [{\citenamefont {Rusca}\ \emph
  {et~al.}(2018{\natexlab{a}})\citenamefont {Rusca}, \citenamefont {Boaron},
  \citenamefont {Gr\"{u}nenfelder}, \citenamefont {Martin},\ and\ \citenamefont
  {Zbinden}}]{Rusca2018}%
  \BibitemOpen
  \bibfield  {author} {\bibinfo {author} {\bibfnamefont {D.}~\bibnamefont
  {Rusca}}, \bibinfo {author} {\bibfnamefont {A.}~\bibnamefont {Boaron}},
  \bibinfo {author} {\bibfnamefont {F.}~\bibnamefont {Gr\"{u}nenfelder}},
  \bibinfo {author} {\bibfnamefont {A.}~\bibnamefont {Martin}},\ and\ \bibinfo
  {author} {\bibfnamefont {H.}~\bibnamefont {Zbinden}},\ }\bibfield  {title}
  {\bibinfo {title} {Finite-key analysis for the 1-decoy state {QKD}
  protocol},\ }\href {https://doi.org/10.1063/1.5023340} {\bibfield  {journal}
  {\bibinfo  {journal} {Applied Physics Letters}\ }\textbf {\bibinfo {volume}
  {112}},\ \bibinfo {pages} {171104} (\bibinfo {year}
  {2018}{\natexlab{a}})}\BibitemShut {NoStop}%
\bibitem [{\citenamefont {Rusca}\ \emph
  {et~al.}(2018{\natexlab{b}})\citenamefont {Rusca}, \citenamefont {Boaron},
  \citenamefont {Curty}, \citenamefont {Martin},\ and\ \citenamefont
  {Zbinden}}]{Rusca2018b}%
  \BibitemOpen
  \bibfield  {author} {\bibinfo {author} {\bibfnamefont {D.}~\bibnamefont
  {Rusca}}, \bibinfo {author} {\bibfnamefont {A.}~\bibnamefont {Boaron}},
  \bibinfo {author} {\bibfnamefont {M.}~\bibnamefont {Curty}}, \bibinfo
  {author} {\bibfnamefont {A.}~\bibnamefont {Martin}},\ and\ \bibinfo {author}
  {\bibfnamefont {H.}~\bibnamefont {Zbinden}},\ }\bibfield  {title} {\bibinfo
  {title} {Security proof for a simplified bennett-brassard 1984
  quantum-key-distribution protocol},\ }\bibfield  {journal} {\bibinfo
  {journal} {Physical Review A}\ }\textbf {\bibinfo {volume} {98}},\ \href
  {https://doi.org/10.1103/physreva.98.052336} {10.1103/physreva.98.052336}
  (\bibinfo {year} {2018}{\natexlab{b}})\BibitemShut {NoStop}%
\bibitem [{\citenamefont {Boaron}\ \emph {et~al.}(2018)\citenamefont {Boaron},
  \citenamefont {Korzh}, \citenamefont {Houlmann}, \citenamefont {Boso},
  \citenamefont {Rusca}, \citenamefont {Gray}, \citenamefont {Li},
  \citenamefont {Nolan}, \citenamefont {Martin},\ and\ \citenamefont
  {Zbinden}}]{Boaron2018b}%
  \BibitemOpen
  \bibfield  {author} {\bibinfo {author} {\bibfnamefont {A.}~\bibnamefont
  {Boaron}}, \bibinfo {author} {\bibfnamefont {B.}~\bibnamefont {Korzh}},
  \bibinfo {author} {\bibfnamefont {R.}~\bibnamefont {Houlmann}}, \bibinfo
  {author} {\bibfnamefont {G.}~\bibnamefont {Boso}}, \bibinfo {author}
  {\bibfnamefont {D.}~\bibnamefont {Rusca}}, \bibinfo {author} {\bibfnamefont
  {S.}~\bibnamefont {Gray}}, \bibinfo {author} {\bibfnamefont {M.-J.}\
  \bibnamefont {Li}}, \bibinfo {author} {\bibfnamefont {D.}~\bibnamefont
  {Nolan}}, \bibinfo {author} {\bibfnamefont {A.}~\bibnamefont {Martin}},\ and\
  \bibinfo {author} {\bibfnamefont {H.}~\bibnamefont {Zbinden}},\ }\bibfield
  {title} {\bibinfo {title} {Simple 2.5{\hspace{0.167em}}{GHz} time-bin quantum
  key distribution},\ }\href {https://doi.org/10.1063/1.5027030} {\bibfield
  {journal} {\bibinfo  {journal} {Applied Physics Letters}\ }\textbf {\bibinfo
  {volume} {112}},\ \bibinfo {pages} {171108} (\bibinfo {year}
  {2018})}\BibitemShut {NoStop}%
\bibitem [{\citenamefont {Li}\ \emph {et~al.}(2022)\citenamefont {Li},
  \citenamefont {Zhang}, \citenamefont {Hao~Tan}, \citenamefont {Liao},
  \citenamefont {Huang}, \citenamefont {Li}, \citenamefont {Wang},
  \citenamefont {Mao}, \citenamefont {Yan}, \citenamefont {Li}, \citenamefont
  {Liu}, \citenamefont {Zhang}, \citenamefont {Peng}, \citenamefont {You},
  \citenamefont {Xu},\ and\ \citenamefont {Pan}}]{Li2022}%
  \BibitemOpen
  \bibfield  {author} {\bibinfo {author} {\bibfnamefont {W.}~\bibnamefont
  {Li}}, \bibinfo {author} {\bibfnamefont {L.}~\bibnamefont {Zhang}}, \bibinfo
  {author} {\bibfnamefont {Y.~L.}\ \bibnamefont {Hao~Tan}}, \bibinfo {author}
  {\bibfnamefont {S.-K.}\ \bibnamefont {Liao}}, \bibinfo {author}
  {\bibfnamefont {J.}~\bibnamefont {Huang}}, \bibinfo {author} {\bibfnamefont
  {H.}~\bibnamefont {Li}}, \bibinfo {author} {\bibfnamefont {Z.}~\bibnamefont
  {Wang}}, \bibinfo {author} {\bibfnamefont {H.-K.}\ \bibnamefont {Mao}},
  \bibinfo {author} {\bibfnamefont {B.}~\bibnamefont {Yan}}, \bibinfo {author}
  {\bibfnamefont {Q.}~\bibnamefont {Li}}, \bibinfo {author} {\bibfnamefont
  {Y.}~\bibnamefont {Liu}}, \bibinfo {author} {\bibfnamefont {Q.}~\bibnamefont
  {Zhang}}, \bibinfo {author} {\bibfnamefont {C.-Z.}\ \bibnamefont {Peng}},
  \bibinfo {author} {\bibfnamefont {L.}~\bibnamefont {You}}, \bibinfo {author}
  {\bibfnamefont {F.}~\bibnamefont {Xu}},\ and\ \bibinfo {author}
  {\bibfnamefont {J.-W.}\ \bibnamefont {Pan}},\ }\bibfield  {title} {\bibinfo
  {title} {High-rate quantum key distribution},\ }\href@noop {} {\bibfield
  {journal} {\bibinfo  {journal} {to be published}\ } (\bibinfo {year}
  {2022})}\BibitemShut {NoStop}%
\bibitem [{\citenamefont {Bosshard}\ \emph {et~al.}(2021)\citenamefont
  {Bosshard}, \citenamefont {Christen}, \citenamefont {H{\"a}nggi},\ and\
  \citenamefont {Hofstetter}}]{Bosshard2021}%
  \BibitemOpen
  \bibfield  {author} {\bibinfo {author} {\bibfnamefont {N.}~\bibnamefont
  {Bosshard}}, \bibinfo {author} {\bibfnamefont {R.}~\bibnamefont {Christen}},
  \bibinfo {author} {\bibfnamefont {E.}~\bibnamefont {H{\"a}nggi}},\ and\
  \bibinfo {author} {\bibfnamefont {J.}~\bibnamefont {Hofstetter}},\ }\bibfield
   {title} {\bibinfo {title} {Fast privacy amplification on gpus}\ }\href
  {https://doi.org/10.5281/zenodo.4551775} {10.5281/zenodo.4551775} (\bibinfo
  {year} {2021}),\ \bibinfo {note} {poster presentation at the 24th Annual
  Conference on Quantum Information Processing (QIP 2021), Online}\BibitemShut
  {NoStop}%
\bibitem [{\citenamefont {Elkouss}\ \emph {et~al.}(2010)\citenamefont
  {Elkouss}, \citenamefont {Martinez-Mateo},\ and\ \citenamefont
  {Martin}}]{Elkouss2010}%
  \BibitemOpen
  \bibfield  {author} {\bibinfo {author} {\bibfnamefont {D.}~\bibnamefont
  {Elkouss}}, \bibinfo {author} {\bibfnamefont {J.}~\bibnamefont
  {Martinez-Mateo}},\ and\ \bibinfo {author} {\bibfnamefont {V.}~\bibnamefont
  {Martin}},\ }\bibfield  {title} {\bibinfo {title} {Information reconciliation
  for quantum key distribution},\ }\href@noop {} {\  (\bibinfo {year}
  {2010})},\ \Eprint {https://arxiv.org/abs/arXiv:1007.1616} {arXiv:1007.1616}
  \BibitemShut {NoStop}%
\bibitem [{\citenamefont {Kiktenko}\ \emph {et~al.}(2017)\citenamefont
  {Kiktenko}, \citenamefont {Trushechkin}, \citenamefont {Lim}, \citenamefont
  {Kurochkin},\ and\ \citenamefont {Fedorov}}]{Kiktenko2017}%
  \BibitemOpen
  \bibfield  {author} {\bibinfo {author} {\bibfnamefont {E.~O.}\ \bibnamefont
  {Kiktenko}}, \bibinfo {author} {\bibfnamefont {A.~S.}\ \bibnamefont
  {Trushechkin}}, \bibinfo {author} {\bibfnamefont {C.~C.~W.}\ \bibnamefont
  {Lim}}, \bibinfo {author} {\bibfnamefont {Y.~V.}\ \bibnamefont {Kurochkin}},\
  and\ \bibinfo {author} {\bibfnamefont {A.~K.}\ \bibnamefont {Fedorov}},\
  }\bibfield  {title} {\bibinfo {title} {Symmetric blind information
  reconciliation for quantum key distribution},\ }\href
  {https://doi.org/10.1103/PhysRevApplied.8.044017} {\bibfield  {journal}
  {\bibinfo  {journal} {Phys. Rev. Applied}\ }\textbf {\bibinfo {volume} {8}},\
  \bibinfo {pages} {044017} (\bibinfo {year} {2017})}\BibitemShut {NoStop}%
\bibitem [{\citenamefont {Mao}\ \emph {et~al.}(2022)\citenamefont {Mao},
  \citenamefont {Li}, \citenamefont {Hao}, \citenamefont {Abd-El-Atty},\ and\
  \citenamefont {Iliyasu}}]{Mao2022}%
  \BibitemOpen
  \bibfield  {author} {\bibinfo {author} {\bibfnamefont {H.-K.}\ \bibnamefont
  {Mao}}, \bibinfo {author} {\bibfnamefont {Q.}~\bibnamefont {Li}}, \bibinfo
  {author} {\bibfnamefont {P.-L.}\ \bibnamefont {Hao}}, \bibinfo {author}
  {\bibfnamefont {B.}~\bibnamefont {Abd-El-Atty}},\ and\ \bibinfo {author}
  {\bibfnamefont {A.~M.}\ \bibnamefont {Iliyasu}},\ }\bibfield  {title}
  {\bibinfo {title} {High performance reconciliation for practical quantum key
  distribution systems},\ }\href {https://doi.org/10.1007/s11082-021-03489-4}
  {\bibfield  {journal} {\bibinfo  {journal} {Optical and Quantum Electronics}\
  }\textbf {\bibinfo {volume} {54}},\ \bibinfo {pages} {1} (\bibinfo {year}
  {2022})}\BibitemShut {NoStop}%
\end{thebibliography}%
\end{document}